\documentstyle[preprint,prabib,aps]{revtex} 

\begin{document}
\title{Wormholes and Flux Tubes in 5D Kaluza-Klein Theory}
\author{V. Dzhunushaliev 
\thanks{E-Mail Address : dzhun@freenet.bishkek.su}}
\address{Dept. of Phys. VCU, Richmond, VA 23284-2000, USA
and Theor. Phys. Dept. KSNU, 720024, Bishkek, Kyrgyzstan}
\author{D. Singleton 
\thanks{E-Mail Address : das3y@maxwell.phys.csufresno.edu}} 
\address{Dept. of Phys., 2345 East San Ramon Ave. M/S 37
Fresno, CA 93740-8031} 
\date{\today}
\maketitle

\begin{abstract}

In this paper spherically symmetric  
solutions to 5D Kaluza-Klein theory, with ``electric'' and/or 
``magnetic'' fields are investigated. It is shown that the global 
structure of the spacetime depends on the relation between the 
``electrical''  and ``magnetic'' Kaluza-Klein fields. For small 
``magnetic'' field we find a wormhole-like solution.
As the strength of the
``magnetic'' field is increased relative to the strength of
the ``electrical''  field, the wormhole-like solution
evolves into a finite or infinite flux tube depending on the
strengths of the two fields. 
For the large ``electric'' field case we conjecture that 
this solution can be considered as the mouth of a wormhole, 
with the $G_{55}$, $G_{5t}$ and $G_{5\varphi}$ 
components of the metric acting
as the source of the exotic matter necessary for the formation 
of the wormhole's mouth. For the large ``magnetic'' field case
a 5D flux tube forms, which is similar to the flux tube  
between two monopoles in Type-II superconductors, or 
the hypothesized color field flux tube between
two quarks in the QCD vacuum.
\end{abstract}

\section{Introduction} 

Spherically symmetric metrics in multidimensional 
(MD) gravity can describe black holes and wormholes 
(WH) (see for, example, \cite{mor}-\cite{vis}). Usually
these papers investigate metrics without off-diagonal  
components. However these components 
of the MD metric can play an important role as
a result of the following theorem \cite{sal}-\cite{per}: 
\par
Let $G$ be the group fibre of the principal bundle. Then 
there is a one-to-one correspondence between the $G$-invariant 
metrics on the  total  space ${\cal X}$
and the triples $(g_{\mu \nu }, A^{a}_{\mu }, h\gamma _{ab})$. 
Here $g_{\mu \nu }$ is Einstein's pseudo  -
Riemannian metric on the base; $A^{a}_{\mu }$ is the gauge field 
of the group $G$ ( the nondiagonal components of 
the multidimensional metric); $h\gamma _{ab}$  is the 
symmetric metric on the fibre. 
\par 
This theorem suggests that including the off-diagonal 
components of the MD metric is equivalent to including 
gauge fields (U(1), SU(2) or SU(3) gauge fields), and 
a scalar field $\phi(x^\mu)$ which is connected with the 
linear size of the extra dimension. 
These geometrical fields can act as the source of the 
exotic matter 
necessary for the formation of the wormhole's mouth. Such 
solutions were obtained in Refs.
\cite{chodos} \cite{clem} \cite{dzh1} \cite{dzh2}. These
solutions are spherically symmetric WH-like
metrics with finite longitudinal size. The throat of these
WH-like solutions is located
between two surfaces where the reduction from 5D
to 4D spacetime breaks down. These results indicate
that the exotic matter necessary for the formation of the WH can 
appear in \textbf{\textit{vacuum multidimensional 
gravity}} from the off-diagonal elements of the metric
(the gauge fields) and from the $G_{55}$ component of the metric
(the scalar field), rather than coming from some externally
given exotic matter. One possible application of this
5D wormhole is to ``sew'' two
Reissner-Nordstrom solutions on to the
two surfaces of the 5D WH solution where the dimensional
reduction from 5D to 4D breaks down. In this
manner one obtains two asymptotically flat 4D regions
with electric flux, which are connected by a 5D
WH throat \cite{dzh3}. The splitting off or compactification
of the extra dimensions is taken to occur at the 
surfaces where the two 4D 
Reissner-Nordstrom solutions are connected to the 5D WH throat.
This composite, asymptotically flat WH has regions
with both compactified extra dimensions (the two exterior
regions of the 4D Reissner-Nordstrom solutions) and with
noncompactified extra dimensions (the 5D throat or bridge which
connects the two 4D solutions). The 5D region of this composite WH 
has a strong gravitational field. 
\par 
In Refs. \cite{dzh1}, \cite{dzh2} a MD metric with
only ``electric'' fields was investigated. In Ref \cite{dzh4}
a MD metric with  ``magnetic'' field = ``electrical'' field  
was investigated.
In this paper we investigate the consequence of having
both ``electric'' and  ``magnetic'' Kaluza-Klein fields of
varying relative strengths. We will consider 5D 
Kaluza-Klein theory as gravity on the principal 
bundle with U(1) fibre and 4D space as the base of this bundle 
\cite{dzh2}. 

\section{Initial equations}

For our spherically symmetric 5D metric we take 
\begin{eqnarray} 
ds^2 &=& e^{2\nu (r)}dt^{2} - r_0^2e^{2\psi (r) - 2\nu (r)}
\left [d\chi +  \omega (r)dt + n\cos \theta d\varphi \right ]^2  
\nonumber \\
&-& dr^{2} - a(r)(d\theta ^{2} + 
\sin ^{2}\theta  d\varphi ^2),
\label{1}
\end{eqnarray}
where $\chi $ is the 5$^{th}$ extra coordinate; 
$r,\theta ,\varphi$ are $3D$  spherical-polar coordinates; 
$n$ is integer; $r \in \{ -R_0 , +R_0 \}$ 
($R_0$ may be equal to $\infty$). We require that all 
functions $\nu (r), \psi(r)$ and $a(r)$ should be even 
functions of $r$ and hence 
$\nu'(0)=\psi'(0)=a'(0)=0$. According to above-mentioned 
theorem $\omega (r)$ is the $t$-component of the electromagnetic 
potential and $(n\cos\theta)$ is the $\varphi$-component. 
This means that we have radial Kaluza-Klein 
``electrical'' and ``magnetic'' fields. 
\par 
Substituting this ansatz into the 5D Einstein vacuum equations 
\begin{equation} 
R_{AB} - \frac{1}{2}G_{AB} R = 0, 
\label{2} 
\end{equation} 
(where $A,B = 0,1,2,3,4$) gives us (using a $REDUCE$ package
for symbolic calculations): 
\begin{eqnarray} 
\nu '' + \nu'\psi' + \frac{a'\nu'}{a} - 
\frac{1}{2} r_0^2 \omega '^2e^{2\psi - 4\nu} = 0,
\label{3}\\
\omega '' - 4\nu'\omega' + 3\omega '\psi ' + 
\frac{a'\omega '}{a} = 0, 
\label{4}\\ 
\frac{a''}{a} + \frac{a'\psi '}{a} - \frac{2}{a} + 
\frac{Q^2}{a^2}e^{2\psi - 2\nu} = 0, 
\label{5}\\ 
\psi '' + {\psi '}^2 + \frac{a'\psi '}{a} - 
\frac{Q^2}{2a^2}e^{2\psi - 2\nu} = 0, 
\label{6}\\ 
\nu '^2 - \nu '\psi ' - \frac{a'\psi '}{a} + 
\frac{1}{a} - \frac{a'^2}{4a^2} - 
\frac{1}{4}r_0^2\omega '^2 e^{2\psi - 4\nu} - 
\frac{Q^2}{4a^2} e^{2\psi - 2\nu} = 0 
\label{7} 
\end{eqnarray} 
here the Kaluza-Klein ``magnetic'' charge is $Q = nr_0$.
The Kaluza-Klein ``electrical'' field can be defined
by multiplying Eq. (\ref{4}) by $4 \pi r_0$ 
and rewriting it in the following way: 
\begin{equation} 
\left( r_0 \omega ' e^{3\psi - 4\nu} 4 \pi a \right)' = 0.
\label{7a} 
\end{equation} 
This can be compared with the normal 4D Gauss's Law 
\begin{equation} 
\left ( E_{4D} S\right )' = 0,
\label{7b}
\end{equation} 
where $E_{4D}$ is 4D electrical field and $S = 4 \pi r^2$ is the 
area of 2-sphere $S^2$. These are five equations for
determining the four ansatz functions ($\nu, \psi ,
a , \omega$). The first four equations (Eqs. (\ref{3} - \ref{6}))
are dynamical equations which determine the ansatz functions, 
while the last equation (Eq. (\ref{7})) contains no new dynamical
information not contained in the first four equations, but
gives some initial conditions related to solving this system
of equations. For the metric given in Eq. (\ref{1})
$r^2$ is replaced by $a(r)$ and the surface area 
is given by $S = 4 \pi a (r)$. Comparing Eq. (\ref{7a})
with Eq. (\ref{7b}) we can identify the 5D Kaluza-Klein
``electric'' field as
\begin{equation}
\label{7c}
E_{KK} = r_0 \omega ' e^{3 \psi - 4 \nu}
\end{equation}
If we integrate Eq. (\ref{7a}) once and let the integration
constant be $4 \pi q$, then from Eq. (\ref{7c}) we
find that $E_{KK} = q / a(r)$ where $q$ can be taken as the
Kaluza-Klein ``electric'' charge.
Finally for the system of equations given in Eqs.
(\ref{3} ) - ( \ref{7} ) we will consider solutions with
the boundary conditions $a(0) = 1 , \psi (0) = \nu (0) = 0$ 
(for numerical calculations we will introduce dimensionless 
function $a(r)\to a(r)/a(0)$ and $x = r/a(0)$).
Using these boundary conditions in Eq. (\ref{7}) and also
in Eq. (\ref{7c}) (which gives $r _0 w' (0) = q$) gives the
following relationship between the Kaluza-Klein ``electric''
and ``magnetic'' charges
\begin{equation}
\label{7d}
1 = {q^2 + Q^2 \over 4a(0)}
\end{equation}
From Eq. (\ref{7d}) it is seen that the charges can be parameterized
as $q = 2 \sqrt{a(0)}\sin \alpha$ and $Q = 2 \sqrt{a(0)}\cos \alpha$.
\par 
We will examine the following different cases: 
\begin{description} 
\item[A)] 
$Q=0$ or $H_{KK} = 0$ , ``magnetic'' field is zero.
\item[B)]
$q=0$ or $E_{KK} = 0$ , ``electrical'' field is zero. 
\item [C)]
$H_{KK} = E_{KK}$, ``electrical'' field equal to ``magnetic'' 
field.  
\item[D)] 
$H_{KK} < E_{KK}$, ``magnetic'' field less 
than ``electrical''. 
\item[E)] 
$H_{KK} > E_{KK}$, ``electrical'' field less 
than ``magnetic''.
\end{description}

\subsection{Switched off ``magnetic'' field.} 

In this case we have the following solution \cite{chodos} \cite{dzh1}: 
\begin{eqnarray}
a & = & r^{2}_{0} + r^{2},
\label{8}\\
e^{2\nu } & = & {2r_{0}\over q}{r^{2}_{0}+r^{2}
\over r^{2}_{0}-r^{2}},
\label{9}\\
\psi &=& 0
\label{9a}\\
\omega & = & 4r_{0}\over q}
{r\over {r^{2}_{0} - r^{2}}.
\label{10}
\end{eqnarray}
This WH-like spacetime has an asymptotical flat metric, 
bounded by two surfaces at $r=\pm r_0$ where the
reduction from 5D to 4D spacetime breaks down. As $r$ moves
away from $0$ the cross-sectional size of the throat, $a(r)$, 
increases. 

A connection can be made between the present solution and 
Wheeler's old proposal of electric charge as a wormhole filled
with electric flux that flows from one mouth to the other --
the ``charge without charge'' model of electric charge. 
In a recent work \cite{dzh3} a model of electric charge 
along these lines was proposed where electric charge is 
modeled as a kind of composite WH with a 
quantum mechanical splitting off of the 5$^{th}$ dimension. 
The 5D WH-like solution of Eqs. (\ref{8}-\ref{10}) 
have two Reissner-Nordstr\"om black holes attached to
it on the surfaces at $\pm r_0$. By considering 4D electrogravity
as a 5D Kaluza-Klein theory in the initial Kaluza formulation 
with $G_{55}=1$ we can join the 5D and Reissner-Nordstr\"om 
solutions at the $r=\pm r_0$ surfaces base to base and 
fibre to fibre.

\subsection{Switched off the ``electrical'' field}

In this case we will simplify by taking $\nu = 0$ in addition to
$\omega =0$ so that the equations reduce to
\begin{eqnarray} 
\frac{y''}{y} + \frac{y'a'}{ya} - \frac{Q^2 y^2}{2a^2} = 0, 
\label{11} \\
\frac{a''}{a} + \frac{y'a'}{ya} - \frac{2}{a} + 
\frac{Q^2 y^2}{a^2} = 0, 
\label{12} \\
\frac{a'y'}{ay} - \frac{1}{a} + \frac{a'^2}{4a^2} + 
\frac{Q^2y^2}{4a^2} = 0 
\label{13} 
\end{eqnarray} 
where $y(r)=\exp{(\psi(r))}$. These are three equations for two
ansatz functions, $\psi (r) , a(r)$. The last equation, Eq. (\ref{13}),
simply repeats information that is already contained in the
first two equations. We solved the system of equations 
(\ref{11}) - (\ref{12}) numerically, using the {\it Mathematica}
package, with the following initial conditions: 
$a(0) = a_0 = 1$, $a'(0) = 0$, $y(0) = 1, y'(0) = 0$, 
(this follows from the fact that 
we can introduce the dimensionless variable 
$x=r/a_0$ and change $a \rightarrow a/a_0$). These conditions
and $\alpha = 0$ fix the dimensionless ``magnetic'' charge as 
$Q =  2$. The result of the numerical calculations 
for $a(r)$ and $y(r)$ are shown in Figs. 1 - 2. We see 
that there is a singularity at two points $x= \pm x_0$ .
Near these singularities we find that the ansatz 
functions have the following asymptotic behaviour: 
\begin{eqnarray} 
y(r) \approx \frac{y_{\infty}}{(r_0 - r)^{1/3}}, 
\label{14} \\
a(r) \approx a_\infty (r_0 - r)^{2/3}, 
\label{15} \\
\frac{Q y_\infty}{a_\infty} = \frac{2}{3}. 
\label{16} 
\end{eqnarray} 
It is interesting to note that the time part of the metric appears 
not to be influenced by the strong gravitational field since 
$G_{tt}(r) = \exp{(2\nu (r))} = 1$. This result is similiar
to what was found in Ref. \cite{gross} \cite{sorkin} where ``magnetic''  
Kaluza-Klein components of the metric were considered. One difference
between the present solutions and the monopole solutions of Ref.
\cite{gross} \cite{sorkin}, is that the monopole solutions had
only coordinate singularities, while $r=\pm r_0$ are real
singularities for the present solution. This can be seen by 
calculating the invariant $R_{AB} R^{AB}$ and using the asymptotic
form for $y(r), a(r)$ given in Eqs. (\ref{14}) - (\ref{16}). This was
done using a $REDUCE$ symbolic calculation package with the result
\begin{equation}
R_{AB} R^{AB} \propto {1 \over (r_0 -r)^2}
\end{equation}
\par 
The cross sectional view of the ansatz function $a(r)$ of
this solution is seen in  Fig. 1. From this 
figure we take the singularities at $r=\pm r_0$ as the 
location of two magnetic charges ($\pm Q$) with opposite signs 
and with flux lines of Kaluza-Klein ``magnetic'' 
field going from $+Q$ to $-Q$. It can be shown that 
this spacetime has a finite volume $V$,
by calculating $V=\int\sqrt{-G}d^5v$. Near the singularities 
$r=\pm r_0$ we have: 
\begin{equation} 
\sqrt{-G} = \sqrt{-det (G_{AB})} = 
r_0 a(r) \exp{(\psi (r)}) \sin\theta 
\approx (r_0 - r)^{1/3} \rightarrow 0 
\label{17}
\end{equation} 
Fig.1 is very suggestive the color field flux 
tubes which are conjectured to form between two quarks 
in some pictures of confinement (see for example pg. 548
of Ref. \cite{peskin}).
\par 
In the preceding section the purely ``electric''
solution of Eqs. (\ref{8} - \ref{10}) can lead to
a Wheeler-like model of ``charge without charge''
\cite{dzh3}. Based on the duality between electric
and magnetic charges \cite{jackson} one might naively
expect that a similiar Wheeler type model for ``magnetic 
charge without magnetic charge'' should exist. However, 
from the 5D Kaluza-Klein magnetic solution of Eqs. (\ref{11}-\ref{13})
we find that the cross-section of the solution given by
$a(r)$ in Fig. 1 decreases as $r \rightarrow \pm r_0$, in
contrast to the ``electric'' solution where $a(r)$ increases
as $r \rightarrow \pm r_0$. Furthermore if 
Reissner-Nordstr{\"o}m solutions are attached to the
``electric'' solution as in Ref. \cite{dzh3} then
one has a model of electric charge where the charges
live in an infinite spacetime. In contrast the ``magnetic''
solutions are confined to a finite spacetime with a flux
tube a Kaluza-Klein ``magnetic'' field running between
the charges.  This may provide a reason why free monopoles 
do not appear to exist in nature : they are confined into
monopole-antimonopole pairs in a finite, flux tube-like spacetime
that is similiar to the flux tube confinement picture of
quarks in QCD. 

\subsection{``Magnetic'' field  equal to ``electrical'' field}

In this case $Q=q$ and an exact solution can be given \cite{dzh4}: 
\begin{eqnarray} 
a = \frac{q^2}{2} = const, 
\label{18}\\
e^{\psi} = e^{\nu} = \cosh\frac{r\sqrt{2}}{q},
\label{19}\\
\omega = \frac{\sqrt{2}}{r_0}\sinh\frac{r\sqrt{2}}{q} 
\label{20}
\end{eqnarray} 
Using this solution and Eq. (\ref{7c}) 
we find that the Kaluza-Klein ``electrical'' field is
\begin{equation} 
E_{KK} = \frac{q}{a} = \frac{2}{q} = const.
\label{23} 
\end{equation}
A similiar magnetic flux tube-like solution was discussed in
Ref. \cite{davidson}.
The Kaluza-Klein ``magnetic'' field can be derived as in
Refs. \cite{gross} \cite{sorkin} . The gauge field associated with the
metric in Eq. (\ref{1}) has a $\varphi$ component as
$A_{\varphi} = r_0 n \cos \theta$. The Kaluza-Klein ``magnetic''
field is then found from ${\bf H}_{KK} = {\bf \nabla} \times
{\bf A}$, where the curl is taken using the metric of Eq. (\ref{1})
and the solution of Eqs. (\ref{18} - \ref{20}). The resultant
Kaluza-Klein ``magnetic'' field derived from this has a magnitude
of  
\begin{equation}
H_{KK} = \frac{r_0 n}{a} = \frac{Q}{a} = const. 
\label{24}
\end{equation}
Thus, this solution is \textit{\textbf{an infinite flux tube}} 
with constant Kaluza-Klein ``electrical'' and 
``magnetic'' fields. The direction of both the
``electric'' and ``magnetic'' fields is along the
${\hat {\bf r}}$ direction ({\it i.e.} along the axis
of the flux tube). The sources 
of these Kaluza-Klein fields (5D ``electrical'' 
and ``magnetic'' charges) are located at $\pm \infty$. 
This feature leads us to consider this solution as a kind
of 5D  ``electrical'' and ``magnetic'' dipole. 

\subsection{Intermediate cases} 

We consider two different cases: 
$E_{KK} > H_{KK}$ (or $q > Q$) and $E_{KK} < H_{KK}$
(or $q<Q$). The initial conditions  for both cases are 
taken as : $\psi (0) = \nu (0) = 0$, 
$\psi '(0) = \nu '(0) = 0$ and $a(0) = 1, a'(0) = 0$. These
initial conditions along with a choice of $\alpha$ determine
the charges $q , Q$. The task of numerically solving the
system of four equations, (\ref{3}) - (\ref{6}), for the
ansatz functions can be simplified by noting that Eq. (\ref{4})
can be integrated out as we did in section II. Using 
$E_{KK} = q / a(r)$ and Eq. (\ref{7c}) we find
\begin{equation}
\label{25}
\omega ' = {q \over r_0 a(r)} e^{4 \nu - 3 \psi}
\end{equation}
In this way the $\omega$ equation has been integrated away and
we can replaced the $\omega '$ term in Eq. (\ref{3}) using
Eq. (\ref{25}), thus reducing the original system of four equations
to three.

\subsubsection{$E_{KK} > H_{KK}$}

The result of a numerical calculation for $a(r)$, using the
{\it Mathematica} package, is presented 
in Fig.3 where we have taken $\alpha = \pi / 3$ so that
$q > Q$. The function $e^{\nu (r)}$ is similiar in form to the
function $y(r)$ in Fig. 2, and it has singularites near
$\pm r_0 = \pm 1.24$. As the ``magnetic'' field increases from $0$ 
to $H_{KK}=E_{KK}$ we find the following : First, compared to 
the WH-like solution
of the pure ``electric'' case, the longitudinal distance 
between the surfaces at $\pm r_0$ is stretched as the magnetic field
strength increases; second, the cross-sectional
size of the solution, represented by the function $a(r)$ does not
increase as rapidly as $r \rightarrow \pm r_0$. In the limit
where the ``magnetic'' field equals the ``electrical'' field, 
$H_{KK}=E_{KK}$, the longitudinal length of the solution
goes to $\infty$ and the cross-sectional  size becomes a
constant.

\subsubsection{$E_{KK} < H_{KK}$}

The result of a numerical calculations is presented 
in Fig.4 where we have taken $\alpha = \pi /6$ so that
$q < Q$.  In this case the ``electrical'' field is taken
as decreasing from the $E_{KK}=H_{KK}$ case down to
$E_{KK}=0$. As the ``magnetic'' field strength increases
relative to the ``electric'' field strength we notice
the following evolution of the solution : the infinite
flux tube of the equal field case turns into a finite flux
tube when $E_{KK}$ drops below $H_{KK}$. Also the cross-sectional
size of this case has a maximum at $r=0$ and decreases as
$r \rightarrow \pm r_0$ where singularities occur. We take these
singularities as the locations of the ``electric'' / ``magnetic''
charges. Between the charges there is a flux tube of
Kaluza-Klein ``electric'' and ``magnetic'' fields. The longitudinal 
size of this flux tube (the distance between charges) reaches its
minimum in the limit when there is only a ``magnetic'' field
($E_{KK} = 0$).

\section{Discussion} 

As the relative strengths of the Kaluza-Klein fields are
varied we find that the solutions to the metric in
Eq. (\ref{1}) evolve in a very interesting and suggestive way.
Starting with the case when there is no ``magnetic'' field
this evolution can be sketched as follows :
\begin{enumerate} 
\item 
$H_{KK} = 0$. The solution 
is \textbf{\textit{a WH-like object}} 
located between two surfaces at $\pm r_0$ where the
reduction of 5D to 4D spacetime breaks down. The cross-sectional
size of this solution increases as $r$ goes from $0$
to $\pm r_0$. The throat between the $\pm r_0$ surfaces is
filled with ``electric'' flux.
\item 
$0 < H_{KK} < E_{KK}$. The solution is again
\textbf{\textit{a WH-like object}}. 
The throat between
the surfaces at $\pm r_0$ is filled with both 
``electric'' and ``magnetic'' fields. The longitudinal
distance between the $\pm r_0$ surfaces increases, and
the cross-sectional size does not increase as rapidly
as $r \rightarrow r_0$, compared to the previous case.
\item 
$H_{KK} = E_{KK}$. In this case the solution is 
\textbf{\textit{an infinite flux tube}} filled
with constant ``electrical'' and ``magnetic'' fields, and
with the charges disposed at $\pm \infty$. The cross-sectional
size of this solution is constant ($ a= const.$). Essentially,
as the magnetic field strength is increased one can think that
the two previous solutions are stretched so that the
$\pm r_0$ surfaces are taken to $\pm \infty$ and the cross section
becomes constant.  
\item 
$0 < E_{KK} < H_{KK}$. In this case we have
\textbf{\textit{a finite flux tube}} located 
between two (+) and (-) ``electrical'' and ``magnetic'' 
charges located at $\pm r_0$. Thus the longitudinal 
size of this object is again finite, but now the cross
sectional size decrases as $r \rightarrow r_0$. At
$r = \pm r_0$ this solution has real singularities which
we interpret as the locations of the charges. 
This solution is very similar to the confinement 
mechanism in QCD where two quarks are 
disposed at the ends of a flux tube with color electrical 
and magnetic fields running between the quarks. 
In this connection one can ask if this similarity 
is accidental or if there is some deeper 
connection between 5D Kaluza-Klein gravity and 
QCD ? We note that in Ref. \cite{singl}
some mappings between 4D gravity and non-Abelian 
theory are discussed.
\item 
$E_{KK} = 0$. This solution is again
\textbf{\textit{a finite flux tube}} only with 
``magnetic'' field filling the flux tube. In this
solution the two opposite ``magnetic'' charges
are confined to a spacetime of fixed volume. This
may indicate why single, asymptotic magnetic charges
have never been observed in Nature : they are permanently
confined into monopole-antimonopole pairs of some fixed
volume.
\end{enumerate} 
\par 
The evolution of the solution from a WH-like object, to
an infinite flux tube, to a finite flux tube, as the
relative strengths of the fields is varied, is presented in
Fig.5. This allow us to make two complimentary conclusions :
First, if one takes some Wheeler-like model of electric
charge as in Ref. \cite{dzh3} then it can be seen that
if the magnetic field becomes too strong the WH-like
solution is destroyed and with it the Wheeler-like
model of electric charge.
Second, if one concentrates a sufficently strong electric
field ({\it i.e.} $E_{KK} > H_{KK}$) into some small region 
of spacetime one is led to the science fiction-like 
possibility that one may be able 
\textit{\textbf{to ``open'' the finite
flux tubes into a WH-like configuration}}. This 
conjecture assumes some kind of spacetime foam model
where the vacuum is populated by virtual flux tubes
filled with virtual ``magnetic'' and/or ``electric''
fields.
\par 
Starting from the solutions obtained here we see that in 5D 
gravity there is a distinction between ``electrical'' and 
``magnetic'' Kaluza-Klein fields. This can be contrasted
with the 4D electrogravity Reissner-Nordstr\"om solution 
which is the same for the electrical and magnetic charges. 

\section{Acknowledgements} 

This work has been funded by the
National Research Council under the Collaboration
in Basic Science and Engineering Program.
The mention of trade names or commercial
products does not imply endorsement by the NRC.

\newpage

\centerline{{\bf Figure Captions}}

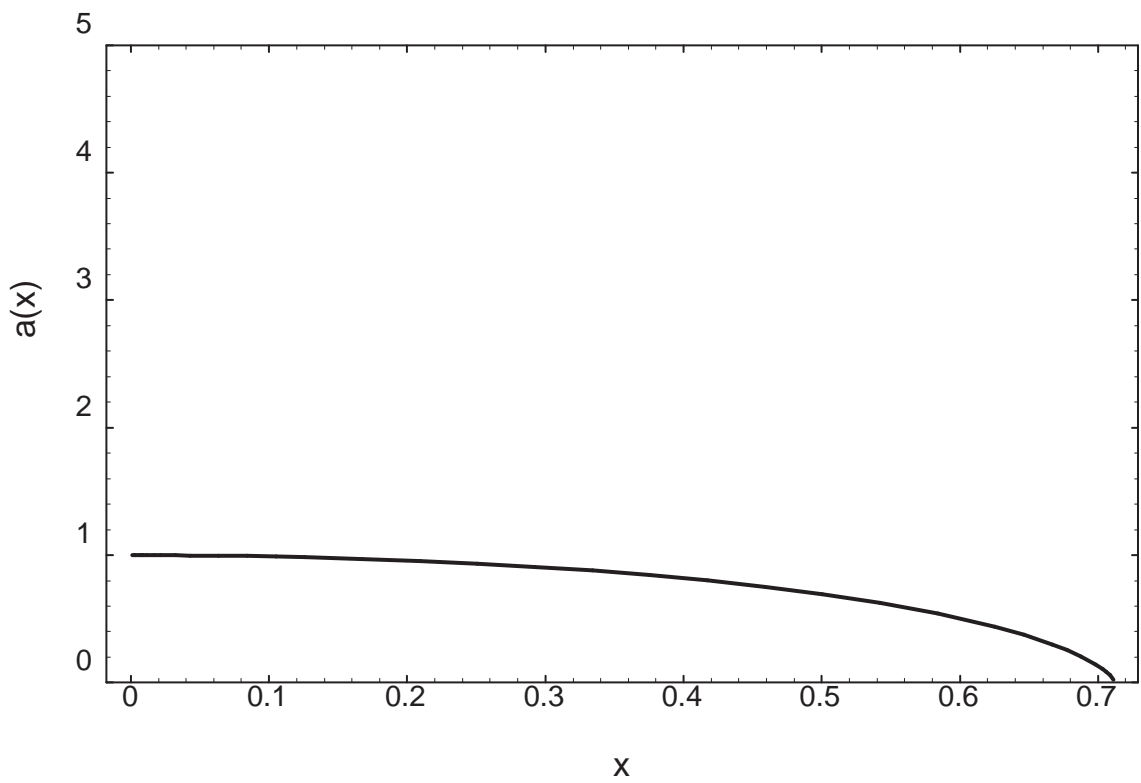
\noindent{{\bf Fig. 1}} Plot of the numerically evaluated ansatz function
$a(x)$ for zero ``electric'' field. The singularities, which
are taken to indicate the locations of the $\pm Q$ ``magnetic''
charges, occur near $+0.71$ and by reflection near $-0.71$. To get
the full picture of the flux tube one should reflect this figure
about the $x$ and $y$ axes.

\vspace{0.2in}

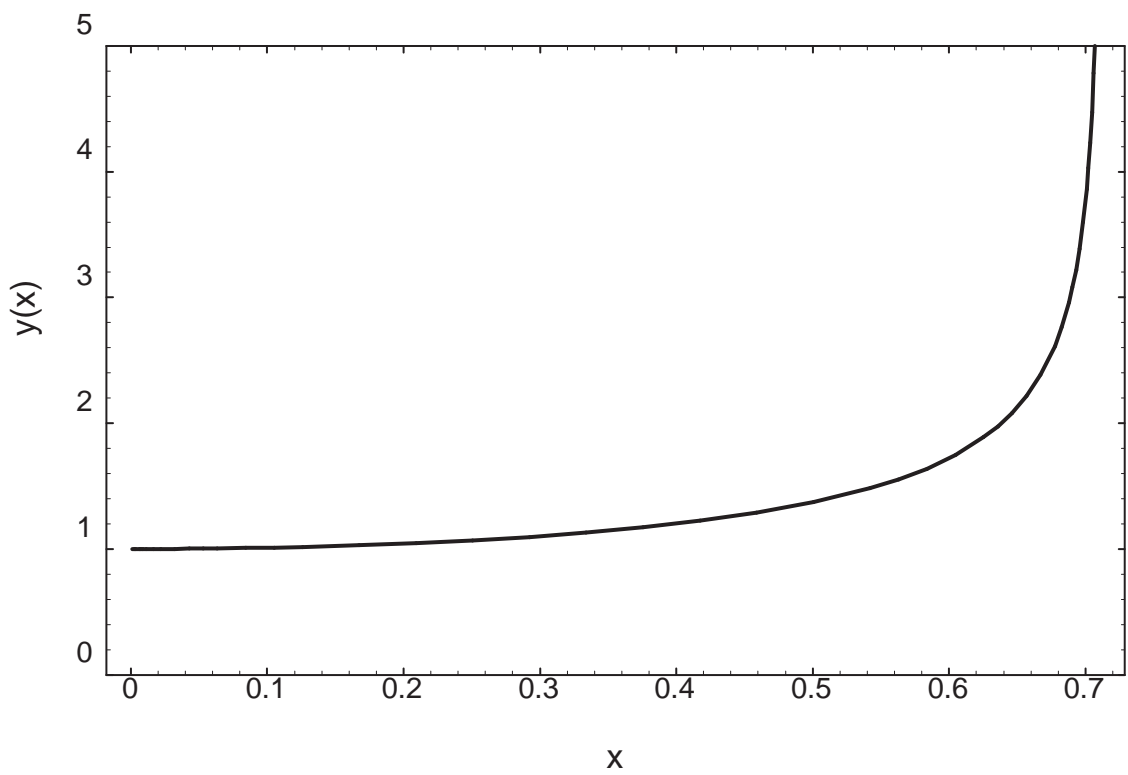
\noindent{{\bf Fig. 2}} Plot of the numerically evaluated ansatz
function $y(x)$ for zero ``electric'' field.

\vspace{0.2in}

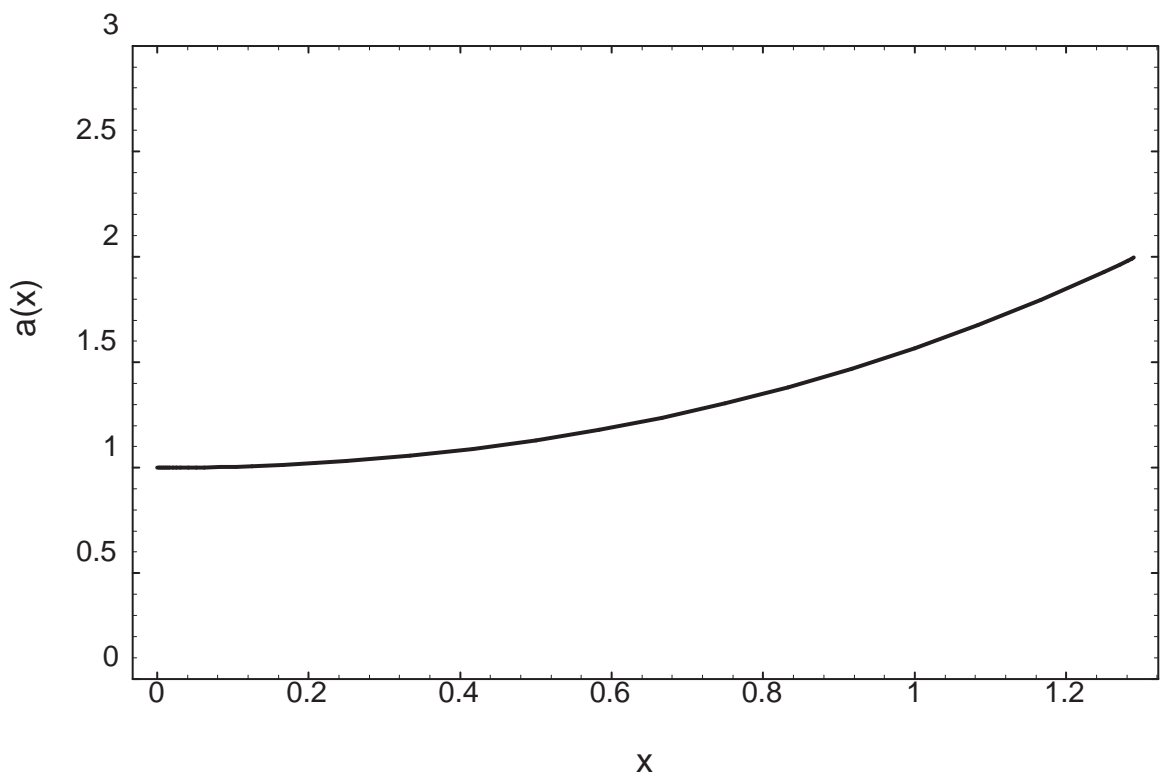
\noindent{{\bf Fig. 3}} Plot of the numerically evaluated ansatz function
$a(r)$ for $\alpha = \pi /3$ ({\it i.e.} the $E_{KK} > H_{KK}$
case). This solution represents a 5D WH-like throat. The ``electric''
and ``magnetic'' charges are taken to be located at the surfaces
near $\pm 1.24$. To get the full picture of the WH one should reflect
this figure about the $x$ and $y$ axes.

\vspace{0.2in}

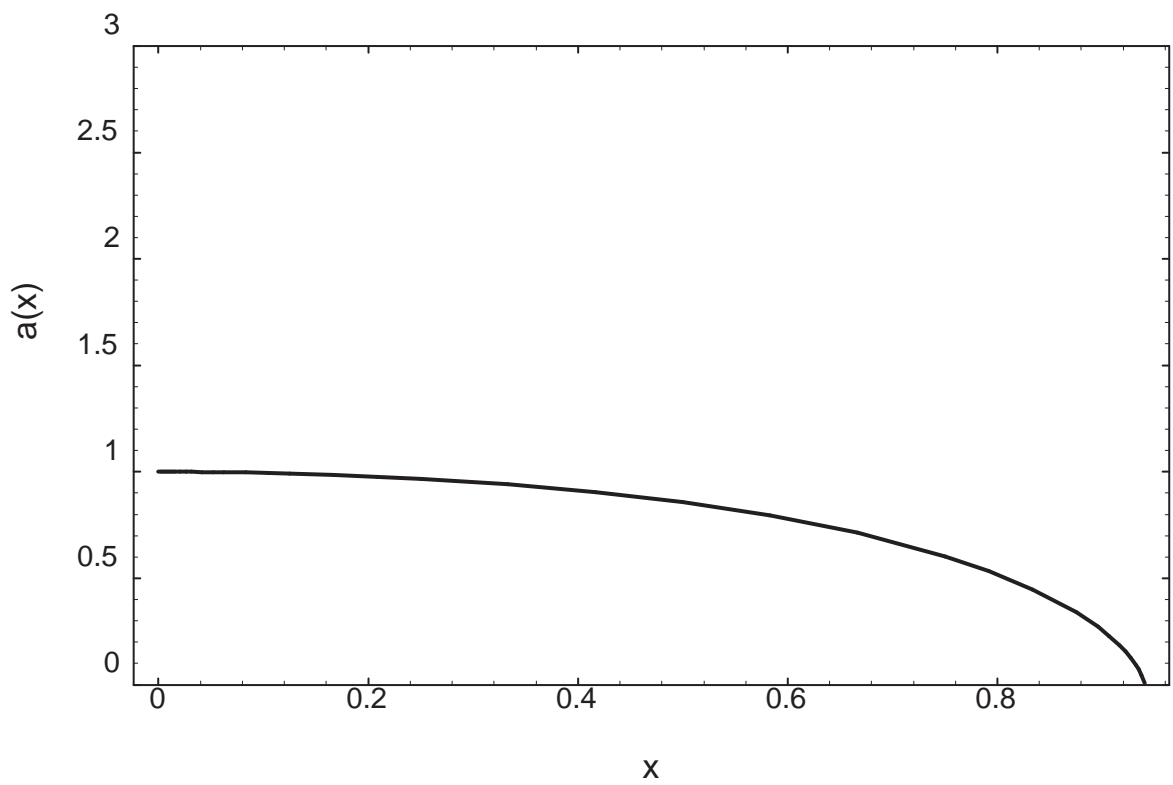
\noindent{{\bf Fig. 4}} Plot of the numerically evaluated ansatz function
$a(r)$ for $\alpha = \pi / 6$ ({\it i.e.} the $H_{KK} > E_{KK}$
case). This figure (after reflection about the $x$ and $y$ axes)
represents a flux tube with the ``electric'' and ``magnetic''
charges located at $\pm 0.87$. Note that relative to the $E_{KK} =0$
case the singularities have moved further apart. This was a
general feature of the solutions ({\it i.e.} the distance between the
singularities increased as the strength of $E_{KK}$ increased).

\vspace{0.2in}

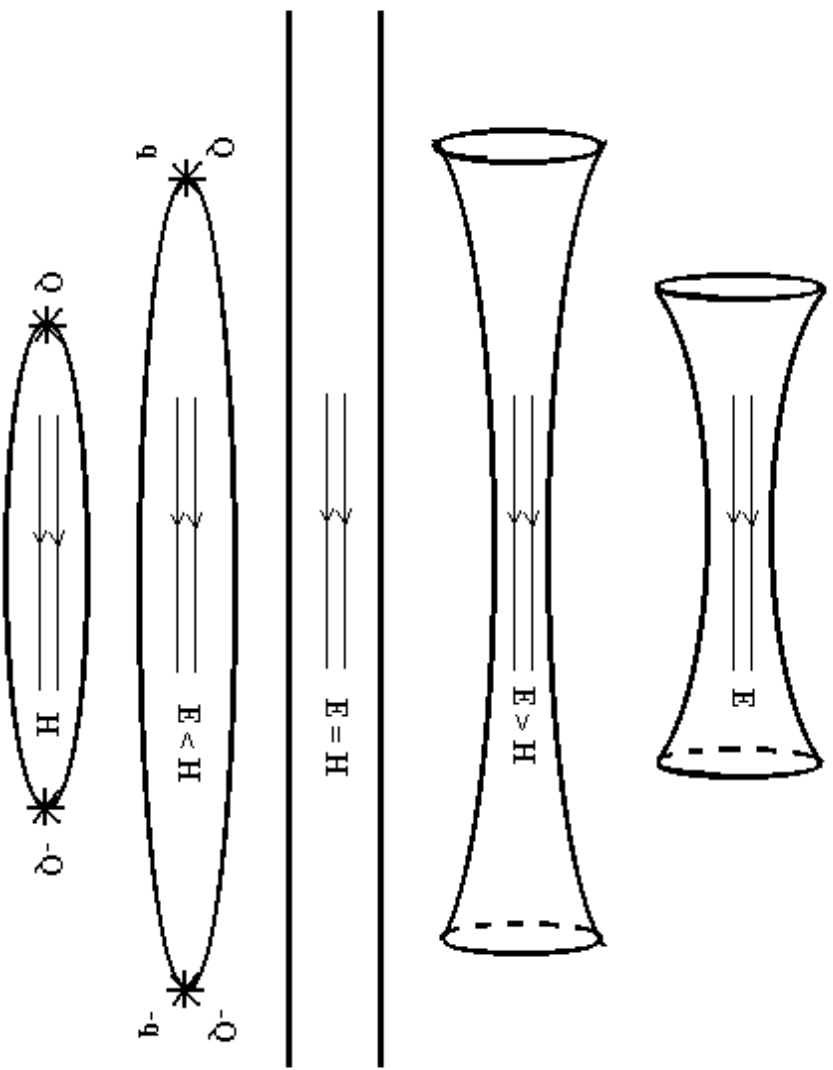
\noindent{{\bf Fig. 5}} The evolution from WH-like like solution to
finite flux tube solution. Starting from the top where
$H_{KK} = 0$ and the solution is WH-like, we find that as
the strength of the ``magnetic'' field is increased the
solution evolves into a stretched WH-like configuration,
an infinite flux tube, a finite flux tube, and finally
for $E_{KK} = 0$ into a finite flux tube with minimal
separation between the charges.

\end{document}